\documentclass[useAMS,usenatbib]{mn2e}
\usepackage{amssymb,amsmath}
\usepackage{graphicx}
\newcommand{\eps}{\epsilon}

\newcommand{\epsi}{\epsilon_{\infty}}

\newcommand{\mcal}{\mathcal}

\newcommand{\be}{\begin{equation}}
\newcommand{\ee}{\end{equation}}

\title[Wisps of the Crab Nebula]
{Observations of ``wisps'' in magnetohydrodynamic simulations of the Crab Nebula }
\author[N. F. Camus et al.]{N. F. Camus,$^{1}$\thanks{E-mail: nfcamus@maths.leeds.ac.uk (NFC)} S. S. Komissarov,
$^{1}$\thanks{E-mail: serguei@maths.leeds.ac.uk (SSK)} N. Bucciantini,$^{2}$ and  P. A. Hughes$^{3}$\\ 
$^{1}$Department of Applied Mathematics, The University of Leeds, Leeds, LS2 9JT \\ 
$^{2}$Astronomy Department, University of California at Berkeley, 601 Campbell Hall, 94720 Berkeley, CA, USA\\
$^{3}$Astronomy Department, University of Michigan, Ann Arbor, MI 48109-1042}
\begin{document}
\date{Received/Accepted}
\maketitle                                    
                                                                                           
\begin{abstract}

In this letter, we describe results of new high-resolution axisymmetric 
relativistic MHD simulations of Pulsar Wind Nebulae. The simulations reveal 
strong breakdown of the equatorial symmetry and highly variable structure of 
the pulsar wind termination shock. The synthetic synchrotron maps, constructed 
using a new more accurate approach, show striking similarity with the well known 
images of the Crab Nebula obtained by \textit{Chandra}, and the 
\textit{Hubble Space Telescope}. In addition to the \textit{jet-torus} structure, 
these maps reproduce the Crab's famous moving \textit{wisps} whose speed and rate 
of production agree with the observations. 
The variability is then analyzed using various statistical methods,  
including the method of structure function and wavelet transform.  
The results point towards the quasi-periodic behaviour with the periods 
of $1.5-3\,$yr and MHD turbulence on scales below $1\,$yr.  
The full account of this study will be presented in a follow up paper.    
\end{abstract}

\begin{keywords}
ISM: supernova remnants -- MHD -- shock waves -- methods: numerical -- 
radiation mechanisms: non-thermal -- relativity -- pulsars: individual: Crab
\end{keywords}

\section{Introduction}
\label{introduction}

The Crab Nebula is the archetypal compact synchrotron nebula, continuously
powered by a magnetized, ultrarelativistic wind from a young, rapidly
rotating neutron star. In order to match the
sub-relativistic expansion of the confining medium (either interstellar medium
(ISM) or supernova remnant (SNR)), the wind must be decelerated. This occurs
at a strong termination shock, established at a radius where the ram pressure
of the pulsar wind matches the pressure of its surrounding \citep{RG74,KC84a}.
This shock causes heating of the wind plasma and presumably particle 
acceleration. As a result a bubble of relativistic particles and magnetic field, 
which is known as a pulsar wind nebula (PWN), 
is formed between the termination shock and the supernova shell
(for general reviews see \citet*{GS06,buc08,KLP07}). 
This way, the spin-down energy lost by the pulsar through its wind becomes 
detectable as non-thermal radiation, mostly synchrotron. 

The steady progress of X-ray astronomy has resulted in the discovery of a
rather peculiar \textit{jet-torus} structure in the inner part of the Crab 
Nebula, not predicted by the classical model of plerions. This
structure had been seen in earlier observations by \cite{bri85} and
\cite{hes95}, but the recent observations using the \textit{Chandra} X-ray
satellite and the \textit{Hubble Space Telescope} 
show the structure and its temporal nature in spectacular detail \citep{wei00,hes02}.  
Similar structures are found in many other PWNe
\citep{wei00,sla04,rom03,cam04,gae01,gae02,got00,hel01,hes95,lu02,pav03}.

The first theoretical models for the formation of both the equatorial torus 
and the polar jets, that combine to form the jet-torus, were presented by
\citet{bog02} and \citet{lyu02}. 
Both these models agree in that the torus arises 
due to the anisotropic distribution of the pulsar wind power, enhanced in the 
equatorial direction, and the jets are formed downstream of the wind termination 
shock.  In particular, \citet{lyu02} proposed that jets are produced   
via the so-called ``tooth-paste'' effect related to 
the hoop stress of the azimuthal magnetic field carried by the wind.   
These theoretical models were confirmed, including the ``tooth-paste'' effect, 
by series of recent numerical simulations \citep{KL03,KL04,ldz04,bog05}, performed using 
newly developed codes for Relativistic MHD (RMHD). 
The RMHD model of PWNe has been recently developed to a high degree
of sophistication, and has proven itself capable of explaining, at least
qualitatively, many properties of PWNe, including their morphology,
polarization, spectral index maps and gamma-ray emission
\citep{buc05,DZ06,vol08,vol09}. 

Although the global structure of Crab's jet-torus appears to be quite persistent 
over long time-scales, it has been known since 1920s that its fine structure shows a short 
time variability both in the vicinity of the termination shock and at larger distances 
\citep{l21,s69}.  The most fascinating variable features are the   
thin synchrotron emitting filaments,  dubbed by \citet{s69} as \textit{wisps}, who 
concluded that they may move outwards with relativistic speeds. However, it is the recent 
high resolution optical and X-ray monitoring observations of the Crab Nebula that have 
enabled us to firmly establish that this is indeed the case  \citep{ttt97,hes02}. 
They have also discovered the jet variability. In particular, the proper motion of the 
jet emission pattern clearly indicates relativistic outflow. Moreover, rather bright and 
variable features appear near the jet base, like the so-called 
\textit{sprite} \citep{hes02}. 
  
\citet{s69} commented that the wisps could be either 1) inhomogeneities advected by the 
flow in the nebula or 2) magnetosonic waves. Later, \citet{GA94} noted that, if ions are part 
of the pulsar wind, then downstream of the termination shock their Larmor radius 
can be comparable to the separation between the brightest wisps of the Crab Nebula, and 
thus the wisps may also reflect the steady kinetic structure of the termination shock as 
mediated by the ions. This idea was recently developed further by \citet{SA04} who have shown 
that the cyclotron instability of ions may lead to the excitation of magnetosonic waves in the 
electron-positron component of the post termination shock flow and thus explain the observed 
variability of the wisps.  \citet{Beg99} proposed an alternative, purely MHD model of 
the wisp origin. He assumed that the flow behind the termination shock consists of a fast  
equatorial zone and a slower moving surrounding zone, and that the
shear between the zones gives 
rise to the Kelvin-Helmholtz instability which causes the equatorial zone to ripple, these 
moving ripples are identified with the dynamic wisps of the Crab Nebula. Recently, 
\citet{buc06} argued that the growth rate of the Kelvin-Helmholtz instability is too 
slow to explain the wisps in this way. On the other hand, the previous RMHD simulations  
of PWN did indicate noticeable variability of the flow and suggested the possibility 
of explaining the wisp production in the MHD framework \citep{KL04,bog05,vol08}.       

In order to explore whether the observed variability of the Crab Nebula can indeed 
be explained within the purely magnetohydrodynamic approach we have carried out new RMHD 
simulations with the highest numerical resolution so far, and developed new sophisticated 
method for the construction of synthetic maps of the synchrotron emission from the simulated PWN. 
In addition, we did not impose the reflective boundary in the  equatorial plane.  
The simulations have revealed the highly unsteady non-linear behaviour of the 
termination shock and the highly intricate fibrous structure of the PWN emission, 
including bright moving wisps. 
Here we briefly describe the key results of this study whereas the   
full account will be given elsewhere.  

\section{Details of numerical simulations}
\label{models}

To perform these simulations we used an improved version of the RMHD scheme
constructed by \cite{SSK99}. In order to reduce numerical diffusion we apply 
parabolic reconstruction instead of the linear one of the original code. Our
procedure, in brief, is to calculate the minmod-averaged first and second
derivatives and use the first three terms of the Taylor expansion for spatial
reconstruction.  

\subsection{Computational grid and initial solution}

We utilise a spherical grid with standard
coordinates $\{r,\theta\}$, where in the $\theta$ direction we adopt a fixed
angular cell size $\Delta \theta$. In the radial direction we adopt a
logarithmic scaling, with radial cell spacing growing as $\Delta r_{i} =
r_{i}\Delta \theta$. As the Courante stability condition requires $\Delta t <
\Delta r_{i}/c$, we split the computational domain into a set of rings such
that the outer radius of each ring is twice its inner radius, and advanced the
solution in each $j$th ring with with its own time step $\Delta t_{j}$, where
$\Delta t_{j+1} = 2\Delta t_{j}$. In order to ensure that the 
termination shock is always fully inside the computational domain and to 
reduce the computational cost, which is dominated by the inner rings, the number of 
rings is variable and is kept to the minimum needed to capture the termination
shock.  If the termination shock moves very close to the inner boundary an 
additional inner ring is activated. When it moves outside of the inner ring, 
this ring is deactivated. Moreover, in order to reduce the computational cost 
further more, we do not update the cells which are fully inside the pulsar wind 
zone. In order to verify the convergence of numerical solutions, in the statistical 
sense, we used four grids with 100, 200, 400, and 800 cells in the $\theta$ direction. 
In this letter, we only present the results obtained with the highest resolution, which 
exceeds by the factor of 4 the resolution of previous simulations.\footnote{ 
The full account of the convergence study will be given in the follow-up paper.}   
In the radial direction the computational domain extends up to the distance 
corresponding to $r = 10\,\mbox{lyr}$ and the simulations are run up 
to the time corresponding to the current age of the Crab Nebula, 
$t \approx 960\,\mbox{yr}$.  

The initial pulsar wind zone is spherical with the radius $r_w=0.5\,\mbox{lyr}$. 
The wind model is exactly the same as in \citet{KL04} with the magnetization 
parameter $\xi=0.4$, the Lorentz factor $W=10$, and purely azimuthal magnetic 
field changing sign at $\theta=\pi/2$ according to the model of dissipation of 
the alternating component of magnetic field of striped wind at the termination shock 
by \citet{Lyu03b}. Initially, 
the region outside of the wind zone is filled with cold radially expanding 
flow with constant density $\rho_e$ and velocity $v\propto r$. The parameters 
of this flow are selected in such a way that its total mass and
kinetic energy  within the radius $r_e=1.47\,\mbox{lyr}$ are $6M_\odot$ and
$10^{51}\,$erg respectively. This leads to the final size of the simulated PWN that is 
consistent with the observations.   
Given the age of the Crab Nebula, it is not expected to have entered the 
reverberation stage and thus the interaction of the supernova shell with ISM 
is not important for the dynamics of the PWN. For this reason we do not introduce the 
ISM zone and continue the supernova shell solution up to the outer boundary of 
the computational domain.      

At $\theta=0$ and $\theta=\pi$ we impose the standard axisymmetric boundary 
conditions using three ghost cells. At the outer radial boundary, $r=10\,\mbox{lyr}$, 
we imposed the non-reflective boundary conditions and at the inner radial boundary 
our boundary conditions describe the supersonic pulsar wind.  
 
Finally, we adopt the equation of state of ideal gas 
with the ratio of specific heats $\gamma = 4/3$.

\subsection{Model of synchrotron emission}
\label{MSE}

The synchrotron particles are assumed to be continuously injected at the 
termination shock with the power law spectrum 
\be 
f(\eps) = A \eps^{-\Gamma}\quad\textrm{for} \quad 
\eps \leqslant \eps_c,
\label{eq:1}
\ee
where the cut-off energy is set to the value $\eps_c=1000$TeV and the power 
index to $\Gamma=2.2$. According to the results of \citet{KC84b} and \citet{vol08} 
this simplified model fits the synchrotron spectrum of the Crab Nebula from 
the optical frequencies to X-rays. In fact, the value of $\Gamma$ is sensitive 
to the model of the nebula flow. Moreover, the origin of radio-emitting electrons 
is unclear and the injection spectrum may have a low energy cut-off close to the 
optical energies. Although these details are important for the spectrum studies, 
they are unlikely to have strong effect on the appearence of the nebula at any given 
frequency between optics and X-rays. The key factors here are the variation of 
magnetic field strength, and the adiabatic and syncrotron energy losses as they has 
strong effect on the relative brigntness of varios features. 

In order to account for the adiabatic losses (or gains), a new dynamic 
variable, the number density of trace species $n$, is introduced 
in the simulations; the variation of $n$ reflects the compression, or 
rarefaction, of the fluid element that carries them around. 
The covariant equation describing its evolution is 
\be
\nabla_{\mu} (n u^{\mu}) = 0,
\ee
where $u^\mu$ is the four-velocity of the flow. In order to deduce 
the total expansion experienced by a particular fluid element downstream 
of the termination shock we also need to know $n_0$, the value of $n$ when this 
element crossed the shock; the volume expansion is then given by the ratio 
$n/n_0$.   This is achieved via treating $n_0$ as another dynamic 
variable which, downstream of the termination shock, satisfies the transport 
equation 
\be
\nabla_{\mu} (n_{0} n u^{\mu}) = 0. 
\label{n0}
\ee
Upstream of the shock we assume that $n_0$ equals to $n$ and that $n$ is  
proportional to the total energy flux, that is  
\be
 n(r,\theta)= N \left(\frac{r_0}{r}\right)^2
 (\sin^2\theta +\frac{1}{\sigma_0}), 
\ee
where $r_0$, $N$, and $\sigma_0$ are constants.  
For simplicity, we assume that, up to a constant coefficient, the parameter $A$
in Eq.\ref{eq:1} equals to $n_0$.  The downstream distribution function of the 
synchrotron electrons can now be expressed as
\be 
f(\eps) = A \left(\frac{n_{0}}{n}\right)^{-\frac{2 + \Gamma}{3}}
\left( 1 - \frac{\eps}{\epsi}\right)^{\Gamma - 2} \eps^{-\Gamma}, 
\quad \eps < \epsi,
\label{eq:10}
\ee
where $\epsi$ is the new cutoff energy due to the synchrotron and adiabatic losses. 
This is another dynamical variable which is evolved in the simulations according to 
the transport equation 
\be
\nabla_{\mu} (\epsilon_{\infty} n^{\frac{2}{3}} u^{\mu}) = 
-\widetilde{c_{2}}B^{2}\epsilon_{\infty}^{2}n^{\frac{2}{3}},
\ee
where $B$ is the magnetic field strength as measured in the fluid frame 
and $\widetilde{c_{2}}=(4e^{4}/9 m^{4} c^{7})$. In the derivation 
we assumed effective pitch-angle scattering of the electrons (and positrons).  
Upstream of the termination shock we set $\epsi=\eps_c$. The corresponding 
synchrotron emissivity, as measured in the observer's frame, is then calculated 
using the approximation 
\be
\mcal{J}_{\nu_{\mbox{\tiny ob}}} \propto  
n_0 \mcal{D}^2 B_{\perp}\left(\frac{n_{0}}{n}\right)^{-\frac{\Gamma + 2}{3}} 
\eps^{1 - \Gamma} \left(1 - \frac{\eps}{\epsi}\right)^{\Gamma - 2},
\ee
where $\nu = c_{1}B_{\perp}\eps^{2}$, $c_{1} = 3e/4\pi m^{3} c^{5}$,
$B_\perp$ is the component of the magnetic field normal to the line 
of sight in the fluid frame, and $\mcal{D}=\nu_{\mbox{\tiny ob}}/\nu$ is the 
Doppler factor.   

\begin{figure}
\includegraphics[width=77mm,height=85mm]{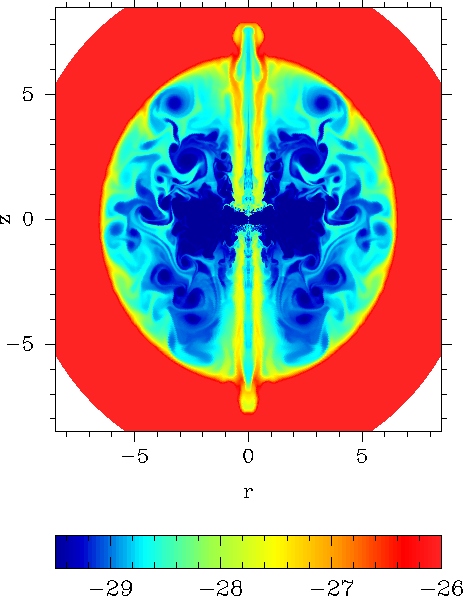}
\caption{The plot shows $log_{10}\rho$ in dimensionless units. The unit of length 
in this and other plots is one light year.} 
\label{fig0}
\end{figure}

\begin{figure*}
\includegraphics[width=58mm,height=64mm]{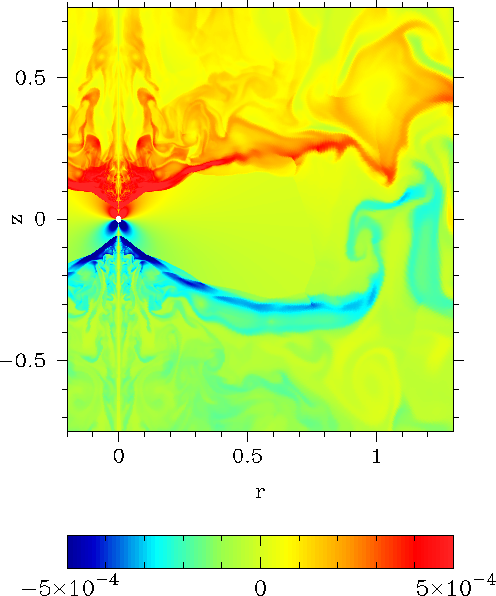}
\includegraphics[width=58mm,height=64mm]{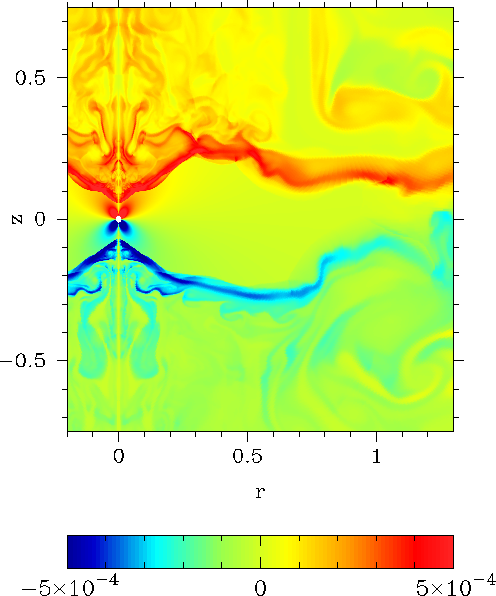}
\includegraphics[width=58mm,height=64mm]{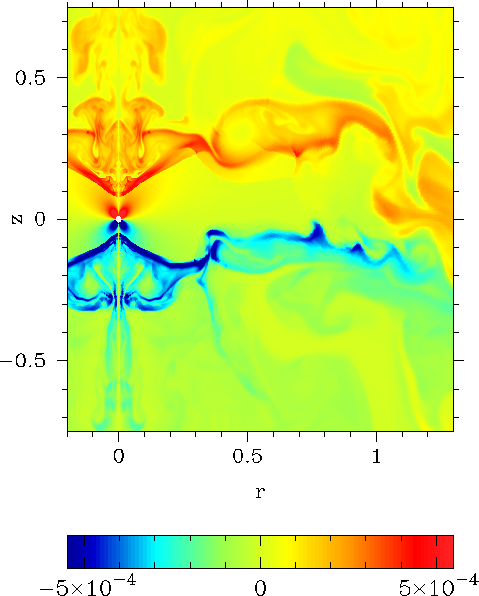}
\caption{ All plots show the azimuthal component of magnetic field in CGS 
units, and clearly illustrate the non-linear 
dynamics of the termination shock. The left panel shows a stage at 
which the termination shock is fully inflated ($t=912.3\,$yr). 
The middle panel shows the northern shock structure being distorted ($t=926.6\,$yr). 
The right panel shows the shock complex to be highly compressed ($t=933.1\,$yr). 
}
\label{fig1}
\end{figure*}
\begin{figure*}
\includegraphics[width=77mm]{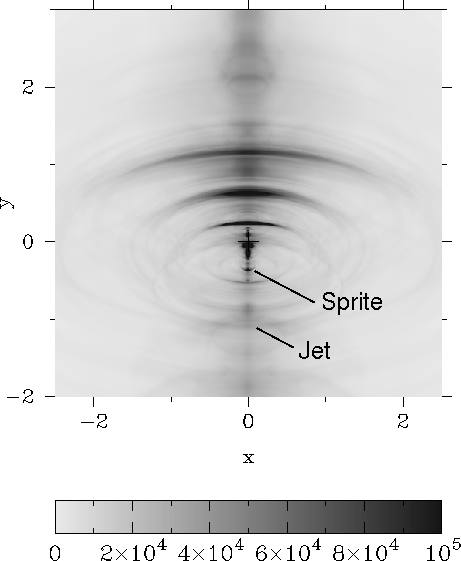}
\includegraphics[width=75.5mm]{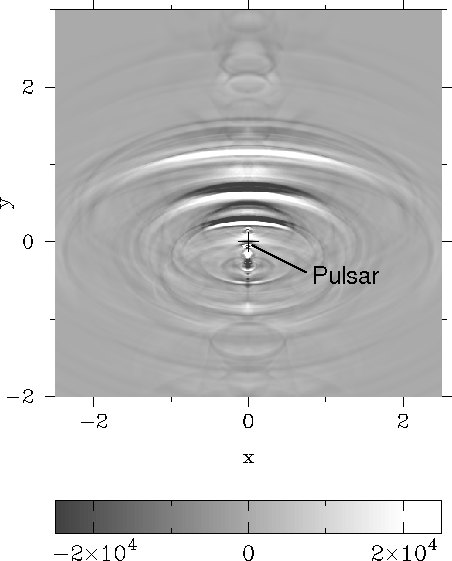}
\caption{The left panel shows the synthetic optical 
synchrotron image of the simulated PWN at time $911.0\,$yr (linear scale).   
The right panel shows the difference between two optical images 
obtained at the same epoch and separated by $\sim 105$ days.}
\label{fig2}
\end{figure*}
\begin{figure*}
\includegraphics[width=56mm,height=60mm]{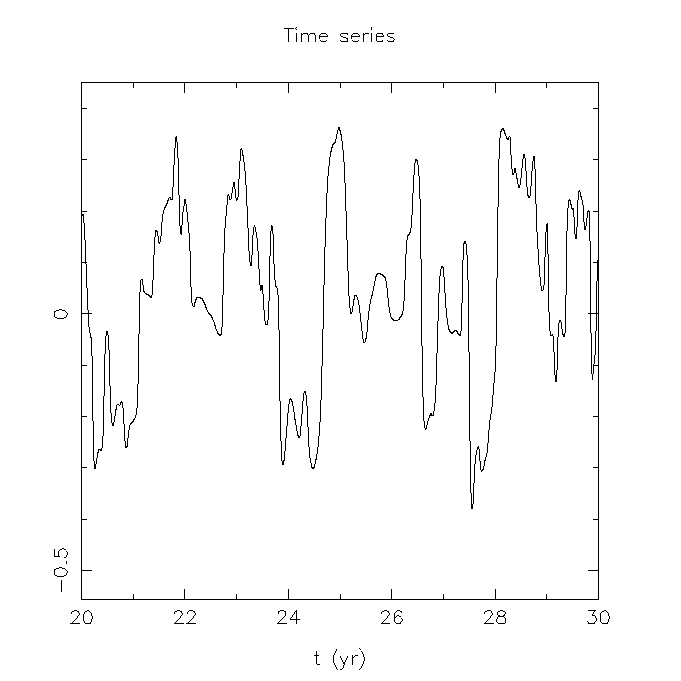}
\includegraphics[width=56mm,height=60mm]{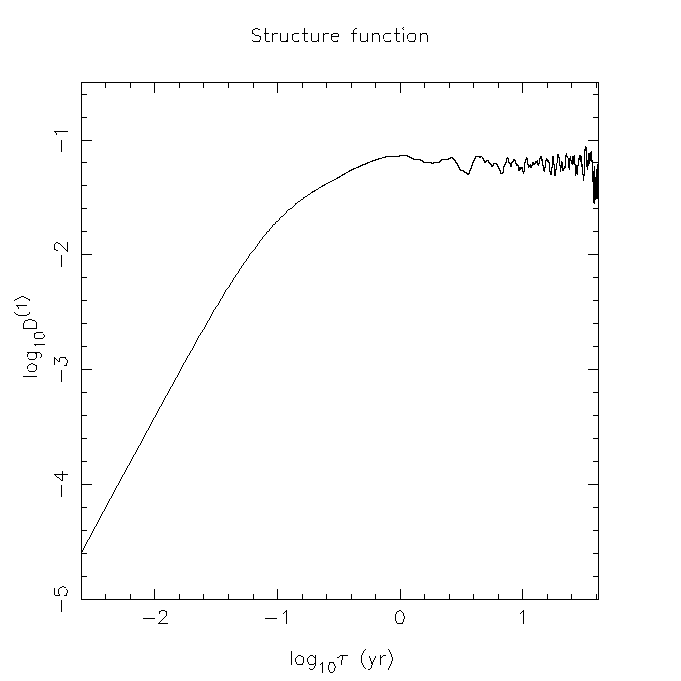}
\includegraphics[width=56mm,height=60mm]{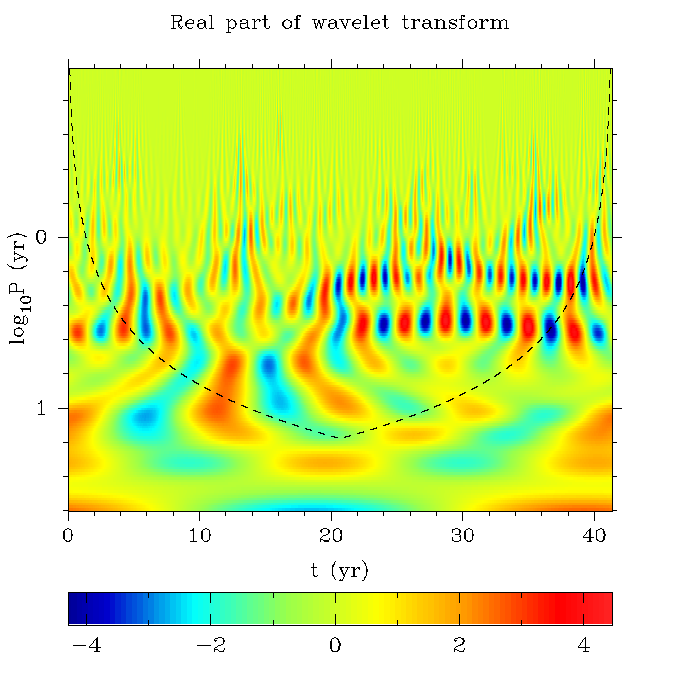}
\caption{ The left panel shows the fragment of time-series data for the 
comoving magnetic field (in arbitrary units) at the reference cell 
with coordinates $r=1.75\,\mbox{lyr}$, 
$\theta = 87^\circ$. Time is measured from the start of sampling. 
The middle panel shows the structure function of this time-series. 
The right panel shows the real part of its  
wavelet transform using the Morlet mother wavelet; 
the dashed line corresponds to the cone of influence.}
\label{fig3}
\end{figure*}

\section{Results and Discussion}

Figure \ref{fig0} shows the solution by the end of the simulation 
run, $t=954\,$yr. Compared to our previous simulations, the most remarkable 
feature of this solution is the absence of the equatorial disk-like outflow 
and the large-scale circulation in the nebula. Instead, we observe a rather 
irregular flow that can be described as a collection of large scale vortexes.  
Since the same behaviour is found 
even in the low resolution runs we conclude that the key factor leading to 
this change in the properties of the PWN flow is the relaxation of the 
equatorial symmetry condition. These vortexes pick up mass from the interface 
between the PWN and the supernova shell and transport it into the nebula's 
interior, creating the intricate fine structure of the density distribution. 
The long ``fingers'' protruding inside the nebula along the polar axis are 
produced by the back-flow of the polar jets contaminated with the entrained mass of 
the supernova shell.  
In the proper three-dimensional case, we do not expect the polar jets to remain 
so straight and collimated as in our simulations. As the result, the backflow 
will be much less pronounced and coherent.

Figure \ref{fig1} illustrates the highly complex dynamics found in the central 
part of the simulated PWN. The structure of the termination shock is highly 
variable due to the non-linear interactions with the flow inside the nebula. 
On one hand, both the size and the shape of the shock change constantly 
due to ``collisions'' with the large amplitude waves and vortexes of the nebula. 
On the other hand, these changes in the shock structure result in highly variable 
outflow from the shock. Thus, the variations of the
termination shock structure and the variability of the flow in the nebula feed on 
each other and at this stage it is quite impossible to distinguish between the 
cause and the effect (this is analogous to the famous ``hen and egg'' problem ).             
This behaviour of the PW termination shock is reminiscent of that of the accretion 
shock in problems involving supersonic accretion \citep[e.g.][]{BMD03,BJRK06,SJFK08} 
and suggests the possibility of common origin. The characteristic time scale of 
the large scale variations is around 1-2 years. This makes sense as the length scale 
of the termination shock is about one light year and the speed of magnetosonic waves 
is close to the speed of light.  

In contrast to the PW, the equatorial symmetry in the PWN is completely broken 
and often the fluid elements blown by the wind into the southern hemi-sphere 
(in Fig.\ref{fig1} they are coloured in blue) end up in the northern hemi-sphere, 
and the other way around. Similarly strong breakdowns of equatorial symmetry 
have been observed in the simulations of stellar collapse \citep[e.g.][]{SJFK08,KB09}. 
This seems to be a general rule for the problems where the shocked 
plasma is confined to a finite or slowly expanding volume.

As expected, and in agreement with previous simulations, the 
predominant motion near the equatorial plane is an outflow. The typical speed of 
this outflow inside the inner $2\,$lyr  is around  $0.6c$.  The outflow is 
highly inhomogeneous with regions of strong magnetic field following  
regions of relatively weak field. Such, strong variations in magnetic field 
strength lead to strong variations of synchrotron emissivity and ultimately 
to the phenomenon of expanding wisps (see Figure \ref{fig2}). 
Further out, but well before reaching the supernova shell the outflow slows down
and mixes with the rest of the PWN. 

Above the equatorial plane a backflow 
with superimposed vortexes can also be seen. This seems to be the reason for 
the contracting wisps occasionally observed in these simulations. Closer to the 
symmetry axis, with $r<0.5\,$lyr, another backflow is seen. Like in the previous 
simulations, this flow originates from the highly-magnetized layers of recently shocked 
plasma of the pulsar wind. The strong magnetic hoop stress in these layers 
stops the outflow and turns it back towards the axis. The resultant axial compression 
drives the polar jets in a fashion reminiscent of the toothpaste flow. 
The axial pinch is rather nonuniform and the flow structure at the jet base is 
highly variable. One can interpret this variability as the development of the 
``sausage'' instability enhanced by the fact that the backflow is already highly  
inhomogeneous and the instability is in the non-linear regime from the start. 
The converging magnetosonic compression waves, which originate in the dynamically 
active region around the equatorial outflow, also contribute to this process.   
In the synthetic synchrotron maps this region often shows  
relatively compact and bright variable features (see Figure \ref{fig2}) which  
look similar to the Crab's  ``sprite'' \citep{hes02}.

The variability of the Crab Nebula emission was investigated by \citet{hes02} 
via subtraction of two images separated by approximately $109$ days. In 
particular, an outward moving wisp appears on the resultant image as a dark 
wisp followed by a bright one.  In Fig.\ref{fig2} we apply the same technique 
to our synthetic optical images. The dynamics of simulated wisps is remarkably 
similar to the that of the real ones.  In agreement with the observations, the typical 
apparent speed of the wisps is $\simeq0.5$c within the inner $\simeq 2\,$lyr 
(In the mapping procedure we ignore the time-delay effect.). 
Further out the apparent speed of the wisps significantly decreases and we often 
observe wisp mergers like those described in \citet{hes02}. Such behaviour agrees 
with the interpretation of wisps as magnetic inhomogeneities advected by a nonuniform
decelerating subsonic flow.

Having observed the wisp production in our RMHD simulations, 
we then quantitatively analysed the nature of the nebula variability in order to 
establish whether the shock dynamics and associated synchrotron emission
show any periodicity, and if not, to determine the statistical nature of the
fluctuations.
A number of time-series were constructed by measuring fluid parameters at the 
point with coordinates $(r,\theta)=(1.75\,\mbox{lyr},87^\circ)$ after each local 
time-step.  
The left panel of Fig.\ref{fig3} shows part of such time series for the 
comoving magnetic field. The middle panel of Fig.\ref{fig3} shows
the structure function of the time-series. One can see that it has a 
plateau at long time lags and a broken power law at short time lags. 
The hard power law at time lags $\tau<0.08\,$yr is clearly a numerical noise as
this time scale corresponds to the numerical resolution of our simulations. 
The softer power law 
at time lags $\tau=0.1-1\,$yr corresponds to well resolved scales and presumably  
captures the MHD turbulence developed in the simulations. 
The transition to plateau at $\tau\simeq1.0\,$yr indicates the characteristic time 
scale for the large scale process that drives the turbulence. The only candidate for 
such a process is the oscillation of the termination shock. This oscillation cannot 
occur on scale below the light crossing time of the shock and this seems to 
explain the characteristic time scale revealed by the structure function.    

In order to investigate this issue further we also applied the method of 
wavelet transform. In the right panel of Fig.\ref{fig3}, the real part of the 
wavelet transform for the Morlet mother wavelet \citep{TC98} is shown. 
One can see clear evidence of quasi-periodic oscillations with the quasi-period 
$P=1.5-3\,$yr. The analysis of time series for other flow parameters gives 
similar results. 

The available variability data for the Crab Nebula do not cover 
sufficiently long time period and are too scarce for rigorous statistical 
analysis. Therefore, we cannot yet make proper comparison between the observations 
and our simulations. The observations by \citet{hes02} revealed two wisps produced 
within one year. This may indicate a somewhat shorter characteristic timescale 
for wisp production compared to other results, but may also be just a local 
variation. Under the circumstances, we conclude that the MHD model explains not only 
the structure of the Crab Nebula but also its variability reasonably well.

\section{Acknowledgments}
NC was supported by the UK Science and Technology Facilities Council (STFC). 
SSK was supported by STFC through the rolling grant ``Theoretical astrophysics 
in Leeds''. NB was supported by NASA through Hubble Fellowship 
grant HST-HF-01193.01-A, awarded by the Space Telescope Science Institute,
which is operated by the Association of Universities for Research in
Astronomy, Inc., for NASA, under contract NAS 5-26555.


\end{document}